\documentclass[%
 reprint,
 amsmath,amssymb,
 aps,
]{revtex4-2}
\usepackage{amsmath}
\usepackage{booktabs}
\usepackage{multirow}
\usepackage{siunitx}
\usepackage{comment}
\usepackage{graphicx}
\usepackage{dcolumn}
\usepackage{bm}

\begin{document}

\preprint{APS/123-QED}

\title{Quantitative Infrared Thermographic Assessment of Hand Cooling Dynamics During Controlled Contact with Metal Plates}

\author{Pengfei Zhu}
 \email{pengfei.zhu.1@ulaval.ca}
 \thanks{Corresponding author}
\affiliation{%
 Department of Electrical and Computer Engineering, Computer Vision and Systems Laboratory (CVSL), Laval University, Quebec, G1V 0A6, Canada
}%

\author{Manyi Zhu}
\affiliation{%
Department of Critical Care Medicine, The Affiliated Chuzhou Hospital of Anhui Medical University (The First People’s Hospital of Chu Zhou), Chu Zhou, China
}%

\author{Guoqing Ren}
\affiliation{%
	Chuzhou Maternal and Child Health Care \& Family Planning Service Center
}%

\author{Stefano Sfarra}
\affiliation{%
	Department of Industrial and Information Engineering and Economics (DIIIE), University of L'Aquila, L'Aquila, I-67100, Italy
}%

\author{Hai Zhang}
\affiliation{%
	Department of Electrical and Computer Engineering, Computer Vision and Systems Laboratory (CVSL), Laval University, Quebec, G1V 0A6, Canada
}%

\author{Clemente Ibarra-Castanedo}
\affiliation{%
	Department of Electrical and Computer Engineering, Computer Vision and Systems Laboratory (CVSL), Laval University, Quebec, G1V 0A6, Canada
}%

\author{Xavier Maldague}
\affiliation{%
	 Department of Electrical and Computer Engineering, Computer Vision and Systems Laboratory (CVSL), Laval University, Quebec, G1V 0A6, Canada
}%

\date{\today}

\begin{abstract}
  This study investigates the spatiotemporal thermal response of human hands during controlled contact cooling using short-wave infrared (SWIR), mid-wave infrared (MWIR), and long-wave infrared (LWIR) thermography. Three participants simultaneously placed one hand on a cooling metal plate and the contralateral hand on a reference plate maintained near room temperature. Temperature evolution was analyzed in five anatomical regions, including the distal finger, proximal finger, vessel-associated region, non-vessel region, and forearm. Quantitative metrics, including temperature variation, bilateral temperature difference, initial cooling rate, and frequency-domain amplitude, were extracted from the thermal image sequences. The results showed that the finger regions exhibited the largest temperature reductions and highest cooling rates, indicating greater sensitivity to thermal stimulation than the dorsal hand and forearm. MWIR and LWIR measurements revealed highly consistent cooling dynamics, while LWIR imaging provided enhanced thermal contrast and sensitivity. Frequency-domain analysis demonstrated that the dominant thermal response was concentrated in the low-frequency range below 0.05 Hz. Furthermore, pixel-wise cooling-rate maps highlighted substantial spatial heterogeneity across the hand surface. Numerical bioheat simulations confirmed that blood perfusion and skin–plate contact conductance are key factors governing the cooling response. These findings demonstrate the potential of dynamic infrared thermography as a non-contact tool for assessing peripheral thermoregulation and vascular function during controlled cooling experiments.
\end{abstract}

\maketitle

\section{Introduction}
Peripheral blood circulation plays a central role in maintaining tissue temperature~\cite{1,2} and supporting physiological thermoregulation in the human extremities~\cite{3,4,5}. The hands are particularly sensitive to environmental temperature because of their large surface-area-to-volume ratio, limited insulating tissue, and dense vascular network~\cite{6,7,8}. Following exposure to a cold environment, cutaneous vasoconstriction reduces peripheral blood flow and heat loss, whereas subsequent vasodilation and blood reperfusion contribute to the recovery of skin temperature. Abnormalities in these regulatory processes may be associated with vascular dysfunction, peripheral neuropathy, repetitive strain injuries, carpal tunnel syndrome, diabetes, and other disorders affecting the microcirculation or autonomic nervous system~\cite{9,10,11,12}. Consequently, quantitative assessment of hand temperature and its response to thermal stimulation may provide useful information about peripheral vascular and thermoregulatory function.

Several established techniques are available for assessing peripheral circulation, including laser Doppler flowmetry, photoplethysmography (PPG), Doppler ultrasound, etc~\cite{13,14,15,16}. These methods are capable of providing direct or indirect measurements of blood-flow dynamics and vascular structure, often with high sensitivity to microvascular changes. However, they are generally limited to single-point measurements or relatively small fields of view, may require physical contact with the skin, and often provide little information about the spatial heterogeneity of thermal and vascular responses across the entire hand. In contrast, infrared thermography does not measure blood flow directly but offers wide-field, non-contact visualization of temperature distributions, enabling simultaneous assessment of multiple anatomical regions and their temporal evolution under physiological or externally induced thermal perturbations.

Infrared thermography provides a non-contact and non-ionizing method for measuring the spatial distribution of skin surface temperature~\cite{17,18,19,20}. In contrast to contact temperature sensors, which record temperature at only a limited number of locations, infrared cameras can simultaneously monitor the entire hand with relatively high spatial and temporal resolution. These characteristics have enabled thermography to be applied to vascular assessment, inflammation detection, musculoskeletal disorders, peripheral neuropathies, and cold-exposure studies~\cite{21,22,23,24}. However, skin temperature is not determined by blood flow alone. It represents the combined effects of tissue conduction, blood perfusion, vascular regulation, metabolic heat generation, and heat exchange with the surrounding environment~\cite{25,26,27,28}. Establishing a quantitative relationship between measured surface temperature and the underlying vascular response therefore remains an important challenge.

Several approaches have been developed to extract blood-flow-related information from infrared measurements. Sagaidachnyi et al. proposed a spectral filtering approach in which low-frequency blood-flow oscillations were treated as sources of thermal waves propagating from the microvasculature toward the skin surface~\cite{29}. By accounting for the frequency-dependent attenuation and phase delay of these thermal components, blood-flow-related signals were reconstructed from thermographic temperature sequences and showed agreement with photoplethysmographic measurements. Similarly, lock-in infrared thermography has been used to enhance the visualization of the vascular tree on the dorsum of the hand~\cite{30}. In that method, periodic changes in vascular perfusion were induced using controlled cuff pressure, and amplitude and phase images were extracted at the stimulation frequency. These studies demonstrate that temporal processing of infrared image sequences can reveal physiological information that is difficult to identify from conventional static thermograms.

The relationship between tissue temperature and circulation has also been investigated using bioheat-transfer models. Conventional models commonly represent the thermal contribution of blood flow through a spatially averaged perfusion term. More detailed vascular–porous-media formulations explicitly account for heat exchange between discrete blood vessels, capillary blood, and surrounding tissue. Wang et al. combined thermographic measurements with such a model to evaluate blood-flow alterations in human hands and feet and demonstrated that regional surface-temperature changes can be associated with altered perfusion and delayed vascular regulation~\cite{31}. These findings also emphasize that absolute temperature alone may be insufficient for characterizing vascular function, because similar temperature patterns may originate from different combinations of tissue properties, vascular structures, boundary conditions, and physiological responses. Dynamic thermal stimulation may therefore provide a more informative assessment by probing the response of the thermoregulatory system rather than observing its resting state alone.

Dynamic thermography has accordingly been introduced in several hand-related applications. Temperature variations during physical exercise have been monitored to characterize the transient thermal response of the hand~\cite{32}, while controlled cooling and subsequent rewarming have been investigated for the assessment of carpal tunnel syndrome~\cite{33,34}. Bargiel et al., for example, cooled the hands before recording their return toward the baseline temperature and observed faster thermal recovery in healthy participants than in patients with carpal tunnel syndrome~\cite{34}. Smartphone-assisted thermography has also recently been evaluated as an accessible tool for identifying regional hand-temperature differences related to carpal tunnel syndrome~\cite{33}. In addition, thermographic asymmetry and localized temperature changes have been investigated in subjects with repetitive strain injuries of the wrist and hand joints~\cite{35}. Although the diagnostic performance reported in these studies varies, they collectively suggest that transient and spatially resolved temperature responses may provide complementary information to static temperature measurements and conventional clinical examination.
Cold stimulation is particularly useful for studying peripheral thermoregulation because it produces an externally controlled perturbation of the thermal and vascular state. During cooling, conductive heat loss and sympathetically mediated vasoconstriction decrease hand temperature, whereas the recovery phase reflects a complex interaction among tissue thermal diffusion, restoration of blood flow, vasomotor regulation, and heat exchange with the environment. Infrared studies of the hand cold pressor test have demonstrated that localized cooling produces not only a substantial decrease in hand temperature but also temporally and spatially distributed responses in the forearm and upper body~\cite{33}. Such observations illustrate that cold exposure induces a systemic thermoregulatory response and that the post-cooling recovery process contains information that cannot be represented by a single minimum-temperature value.
Despite these developments, most previous studies have focused on either static thermal asymmetry, disease classification, spectral extraction of spontaneous blood-flow oscillations, or thermal responses to water immersion and pressure-cuff stimulation. Comparatively less attention has been given to the spatiotemporal response of both hands during paired contact with metal surfaces maintained at different temperatures. Contact cooling differs from water immersion because heat is transferred primarily through a defined solid–skin interface, allowing the cooling temperature, contact duration, and contact area to be more directly controlled~\cite{36,37}. Moreover, comparison with a room-temperature metal plate can help distinguish the effect of cold stimulation from the effects of contact pressure, posture, and conductive heat exchange with the metal surface.

In the present study, infrared thermography was used to investigate the bilateral hand-temperature response of three participants during contact with a cooling metal plate and a second metal plate maintained at room temperature. Rather than relying exclusively on absolute skin temperature, the analysis focuses on the complete temporal response, including the initial temperature distribution, cooling amplitude, regional temperature evolution, and post-stimulation thermal recovery. Spatially resolved temperature curves are extracted from the fingers, palm, and other anatomically relevant regions to examine whether the cold-induced response is homogeneous across the hand. The room-temperature plate is used as a contact-reference condition to account for temperature changes caused by contact with a thermally conductive surface in the absence of strong external cooling.

The objective of this pilot study is to establish a quantitative thermographic framework for characterizing the spatiotemporal response of the human hand to controlled contact cooling. It is hypothesized that the cooled condition will produce not only a larger temperature decrease but also distinct regional and temporal recovery characteristics compared with room-temperature contact. The results are expected to clarify the value of dynamic recovery parameters over conventional static temperature measurements and provide a methodological basis for future investigations involving larger populations and subjects with impaired peripheral circulation or thermoregulatory dysfunction.

\section{Theoretical Framework}
The skin temperature measured by infrared thermography is determined by the combined effects of heat conduction within the tissue, blood perfusion, metabolic heat generation, and heat exchange with the surrounding environment. At the continuum scale, these coupled thermal processes are commonly described by the Pennes bioheat equation~\cite{36},
\begin{equation}
	\rho_t c_t
	\frac{\partial T(\mathbf r,t)}{\partial t}
	=
	\nabla\cdot
	\left(
	k_t\nabla T
	\right)
	+
	\rho_b c_b
	\omega_b
	\left(
	T_b-T
	\right)
	+
	Q_m,
	\label{eq:pennes}
\end{equation}
where $T$ denotes the tissue temperature, $\rho_t$, $c_t$, and $k_t$ are the tissue density, specific heat, and thermal conductivity, respectively, while $\rho_b$, $c_b$, and $\omega_b$ represent the corresponding blood properties and perfusion rate. $Q_m$ is the metabolic heat generation.
In the present work, Eq.~(\ref{eq:pennes}) is not solved numerically. Instead, it provides the physical basis for interpreting the transient temperature evolution observed by infrared thermography during controlled cooling.
When the hand is placed on a metal plate, conductive heat transfer dominates the thermal exchange across the skin--plate interface. The boundary condition can be approximated as
\begin{equation}
	-k_t
	\frac{\partial T}{\partial n}
	=
	h_c
	(T_s-T_p),
	\label{eq:boundary}
\end{equation}
where $T_s$ is the skin temperature, $T_p$ is the plate temperature, and $h_c$ is the effective thermal contact conductance.
For the cold-stimulated hand,
\begin{equation}
	q_{\rm cold}
	=
	h_c
	(T_s-T_{\rm cold}),
\end{equation}
whereas the reference hand satisfies
$
	q_{\rm ref}
	=
	h_c
	(T_s-T_{\rm ref}).
$
Since
$
T_{\rm cold}<T_{\rm ref},
$
the conductive heat flux satisfies
$
q_{\rm cold}>q_{\rm ref},
$
under comparable contact conditions. Consequently, the room-temperature plate serves as a contact-reference condition that compensates for the thermal influence of metal contact itself while isolating the additional cooling produced by the low-temperature plate.
For a semi-infinite medium subjected to a sudden surface-temperature change, the characteristic thermal penetration depth approximately follows
\begin{equation}
	\delta
	\approx
	2\sqrt{\alpha t},
	\label{eq:depth}
\end{equation}
where
$
\alpha
=
\frac{k_t}{\rho_t c_t}
$
is the tissue thermal diffusivity~\cite{39,40,41}.
Equation~(\ref{eq:depth}) indicates that thermal diffusion initially affects only superficial tissues and progressively propagates toward deeper regions with increasing contact time. Consequently, the measured skin temperature reflects both conductive heat transfer and physiological regulation of peripheral blood circulation.
Rather than relying solely on absolute temperature, the transient thermal response is characterized using several quantitative parameters extracted from the infrared image sequences.
The temperature change relative to the initial state is defined as
\begin{equation}
	\Delta T(\mathbf r,t)
	=
	T(\mathbf r,t)-T_0(\mathbf r),
	\label{eq:dT}
\end{equation}
where $T_0$ denotes the initial temperature immediately before cooling.
The cooling amplitude during the observation period is
\begin{equation}
	A_{\rm cool}
	=
	T_0-T_{\min},
	\label{eq:Acool}
\end{equation}
where $T_{\min}$ is the minimum temperature recorded during the experiment.
To quantify the initial cooling behavior, the early-stage cooling rate is obtained by linear fitting,
\begin{equation}
	R_c
	=
	-
	\frac{dT}{dt},
	\label{eq:coolingrate}
\end{equation}
where the derivative is evaluated over the initial approximately linear cooling interval. This parameter characterizes the initial heat extraction efficiency and is less affected by long-term thermal equilibrium.
To eliminate the influence of different baseline temperatures among participants, the normalized temperature variation is defined as
\begin{equation}
	\Theta(\mathbf r,t)
	=
	\frac{T(\mathbf r,t)-T_0(\mathbf r)}
	{T_0(\mathbf r)}.
	\label{eq:normalize}
\end{equation}
The thermal response induced by cold stimulation is evaluated by comparing the cooled hand with the contralateral reference hand,
\begin{equation}
	\Delta T_{\rm cold-ref}
	=
	T_{\rm cold}
	-
	T_{\rm ref}.
	\label{eq:coldref}
\end{equation}
This paired comparison suppresses common environmental effects and emphasizes the additional temperature variation induced by the cooling plate.

Besides the temporal evolution, frequency-domain analysis is employed to investigate the dynamic thermal response.
For each pixel, the discrete Fourier transform of the temperature sequence is calculated as~\cite{42,43,44}
\begin{equation}
	F(f)
	=
	\sum_{n=0}^{N-1}
	T(n)
	\exp
	\left(
	-i2\pi fn/N
	\right),
	\label{eq:fft}
\end{equation}
where $N$ denotes the total number of recorded frames.
The corresponding amplitude spectrum is
\begin{equation}
	A(f)
	=
	|F(f)|.
	\label{eq:amp}
\end{equation}
The amplitude images at different frequencies reveal the spatial distribution of thermal fluctuations occurring over different temporal scales. Compared with raw temperature images, frequency-domain analysis suppresses random temporal noise and enhances slowly varying thermal responses associated with contact cooling.
All quantitative parameters were calculated independently for each participant and each anatomical region of interest (ROI). Cooling amplitudes, temperature differences, and initial cooling rates are reported as the mean values obtained from the selected ROIs. Initial cooling rates were estimated using linear least-squares fitting over the early approximately linear portion of each cooling curve. Unless otherwise stated, comparisons between the cooled and reference hands were performed on paired measurements obtained simultaneously under identical environmental conditions.

\begin{figure*}[t]
	\centering
	\includegraphics[width=\textwidth]{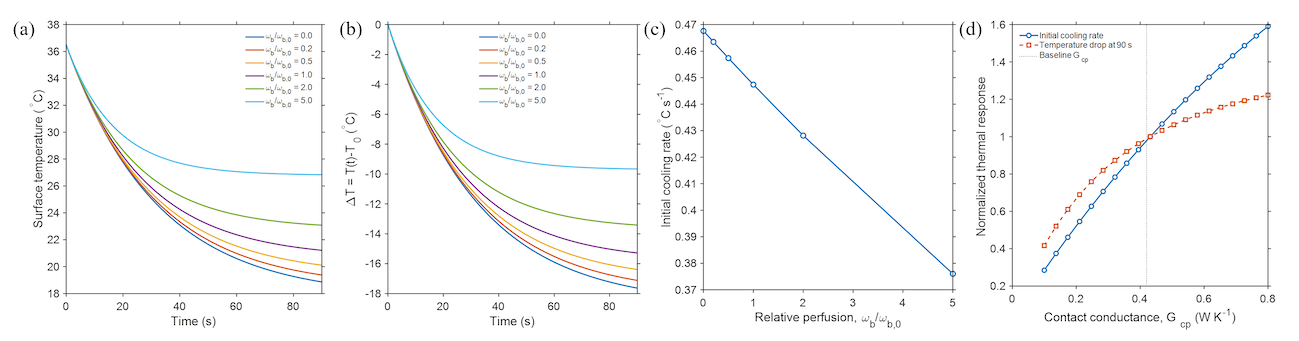}
	\caption{Numerical predictions of hand-surface cooling based on a sustained contact-cooling bioheat model: (a) surface-temperature evolution for different relative blood-perfusion levels; (b) temperature change relative to the initial temperature; (c) initial cooling rate versus relative perfusion; and (d) normalized thermal response as a function of skin–plate contact conductance. The dashed vertical line denotes the baseline contact conductance used in the simulations.}\label{fig1}
\end{figure*}

\begin{figure*}[t]
	\centering
	\includegraphics[width=\textwidth]{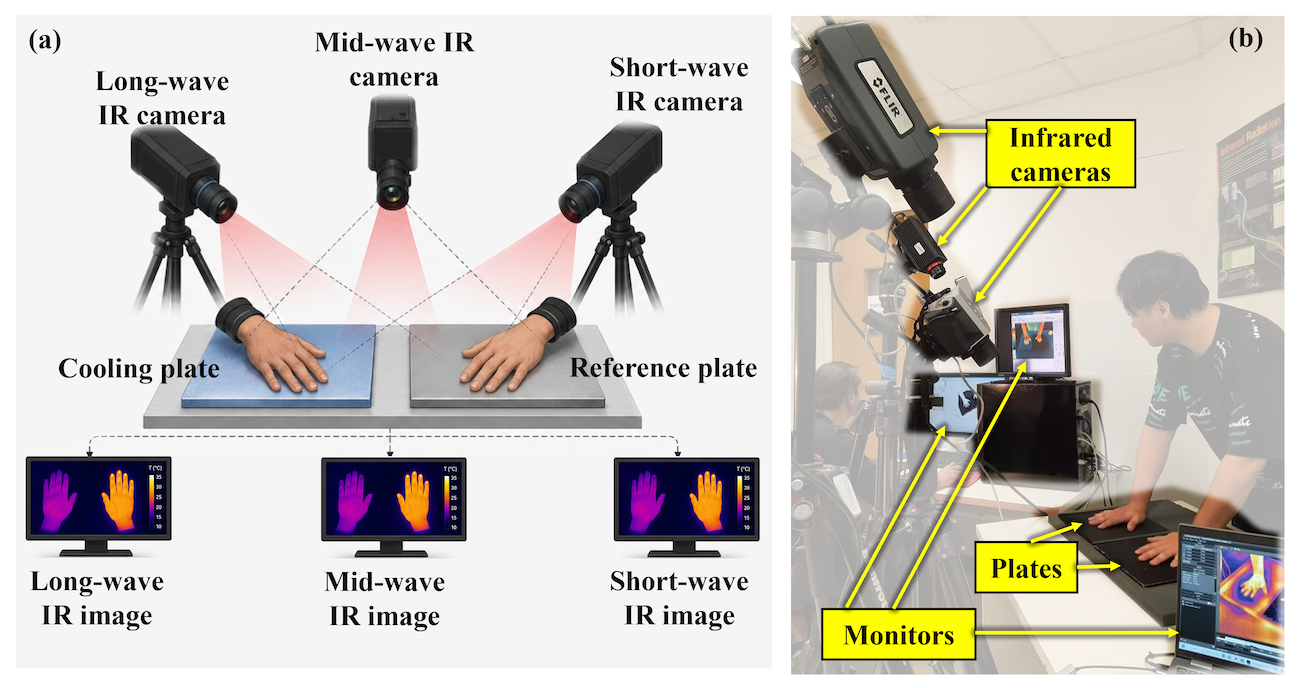}
	\caption{Experimental setup for bilateral hand contact-cooling thermographic measurements. (a) Schematic illustration of the measurement configuration. One hand was placed on a temperature-controlled cooling plate, while the contralateral hand was positioned on a reference metal plate maintained near room temperature. Simultaneous thermal monitoring was performed using long-wave infrared (LWIR), mid-wave infrared (MWIR), and short-wave infrared (SWIR) imaging systems. (b) Photograph of the experimental platform showing the infrared cameras, monitoring system, and metal plates used during data acquisition.}\label{fig2}
\end{figure*}

\begin{figure*}[t]
	\centering
	\includegraphics[width=\textwidth]{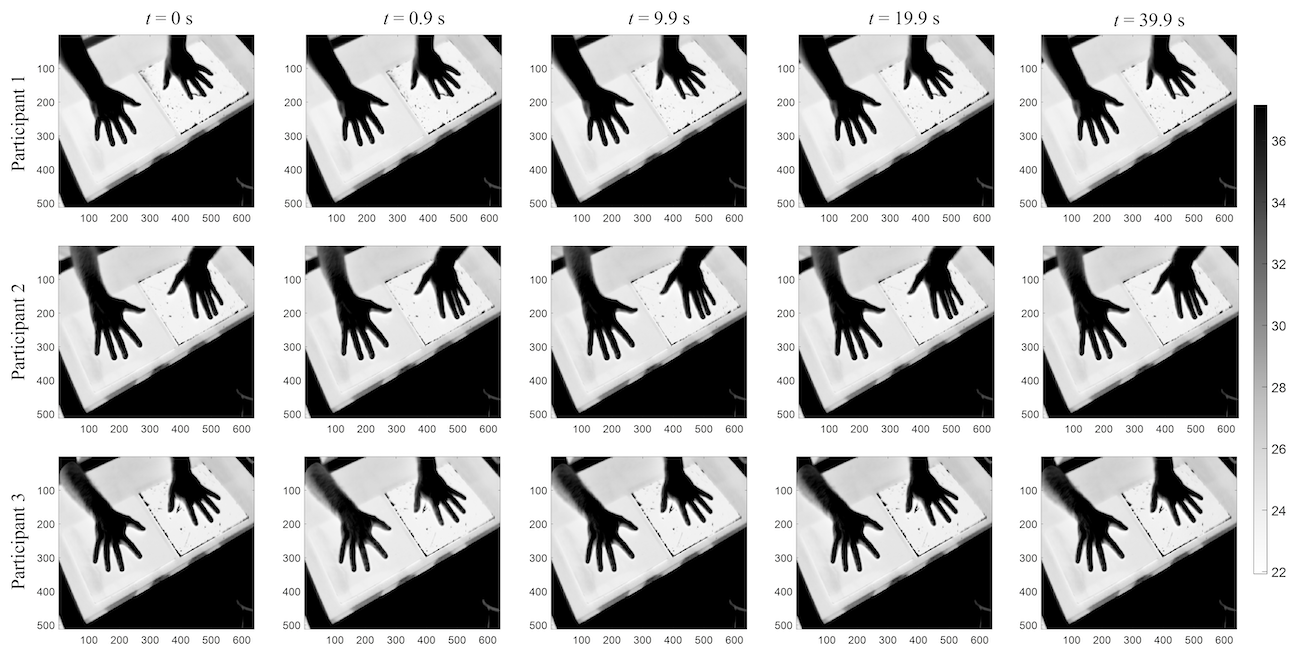}
	\caption{Representative SWIR thermograms acquired during the contact-cooling experiment for three participants. Thermal image sequences recorded at $t=0, 0.9, 9.9, 19.9,$ and 39.9 s are shown for Participants 1–3. During the experiment, one hand was placed on the cooling plate while the contralateral hand rested on the reference plate. Progressive temperature reduction is observed in the cooled hand with increasing contact time, whereas the reference hand exhibits comparatively smaller thermal variations. The color scale represents the measured surface temperature in degrees Celsius.}\label{fig3}
\end{figure*}

\begin{figure*}[t]
	\centering
	\includegraphics[width=\textwidth]{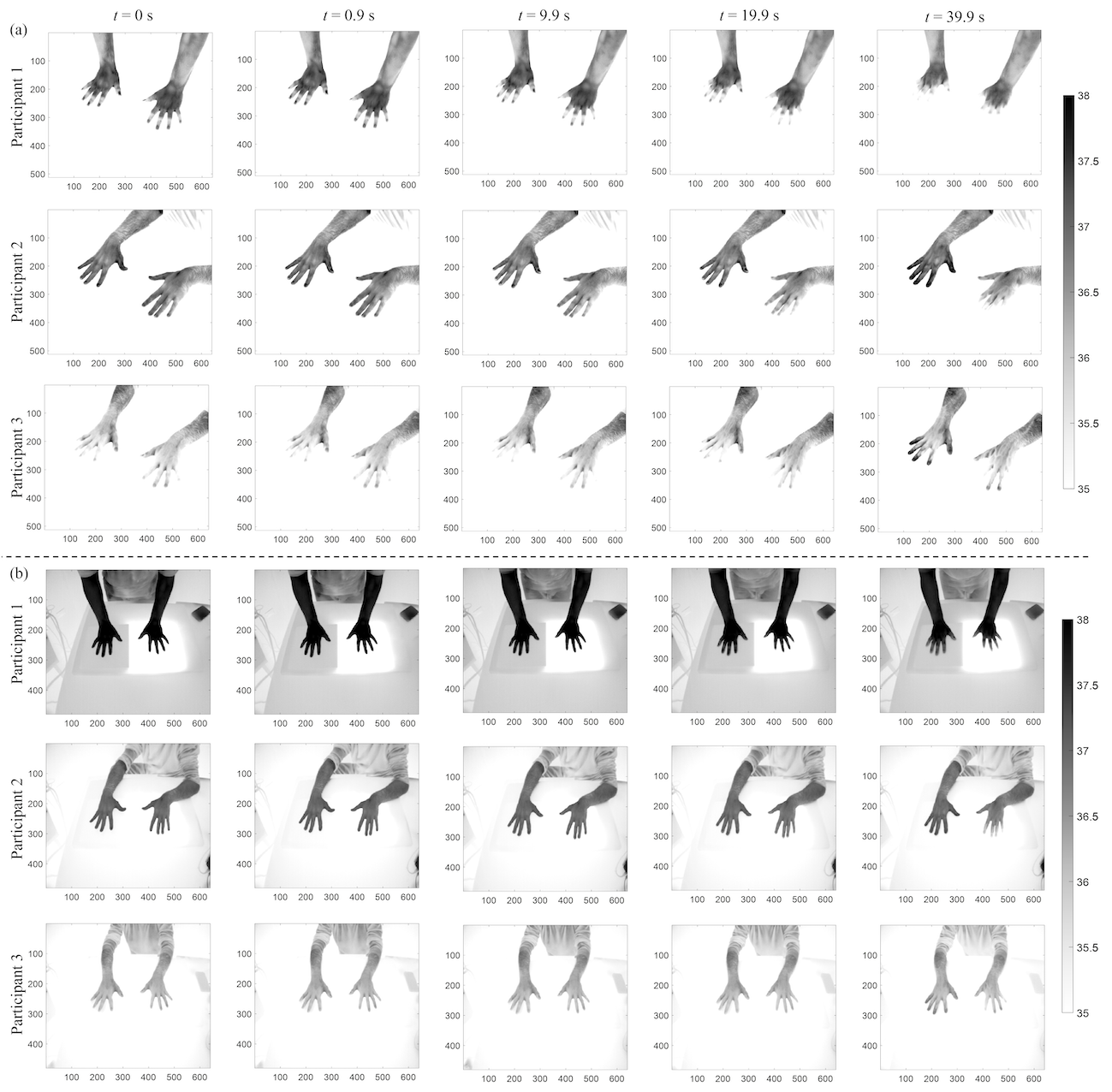}
	\caption{Multi-temporal infrared thermograms acquired during the contact-cooling experiment for three participants. (a) Mid-wave infrared (MWIR) image sequences recorded at $t=0$, 0.9, 9.9, 19.9, and 39.9 s. (b) Long-wave infrared (LWIR) image sequences acquired at the corresponding time instants. During the experiment, one hand was placed on the cooling plate and the contralateral hand on the reference plate. Progressive thermal changes can be observed throughout the cooling period, with the cooled hand exhibiting a more pronounced temperature reduction than the reference hand. The MWIR and LWIR measurements reveal similar overall cooling trends while providing complementary thermal information in different infrared spectral bands. The grayscale bars indicate the measured surface temperature in degrees Celsius.}\label{fig4}
\end{figure*}

\begin{figure*}[t]
	\centering
	\includegraphics[width=\textwidth]{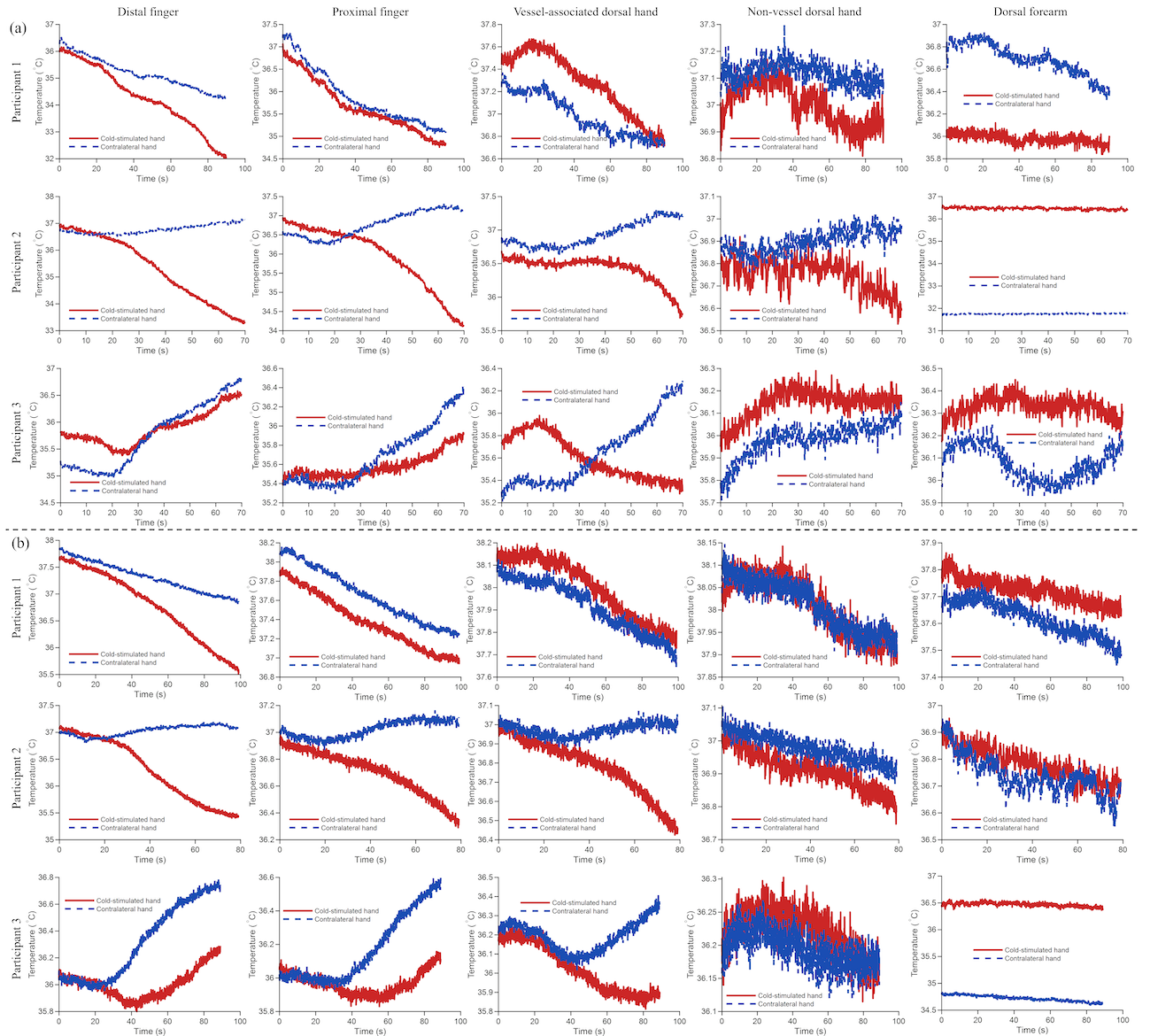}
	\caption{Temporal temperature evolution extracted from five anatomical regions of interest (ROIs) during the contact-cooling experiment. (a) Mid-wave infrared (MWIR) results and (b) long-wave infrared (LWIR) results. The columns correspond to the distal finger, proximal finger, vessel-associated dorsal-hand region, non-vessel dorsal-hand region, and dorsal forearm, respectively. The rows represent Participants 1–3. Solid red curves denote the cold-stimulated hand, whereas dashed blue curves represent the contralateral reference hand. The temperature responses reveal region-dependent cooling behaviors, with the finger regions generally exhibiting larger temperature variations than the dorsal hand and forearm regions. Comparison between the stimulated and contralateral hands highlights the thermal effects induced by contact cooling and the associated spatial heterogeneity of the hand thermoregulatory response.}\label{fig5}
\end{figure*}

\begin{figure*}[t]
	\centering
	\includegraphics[width=\textwidth]{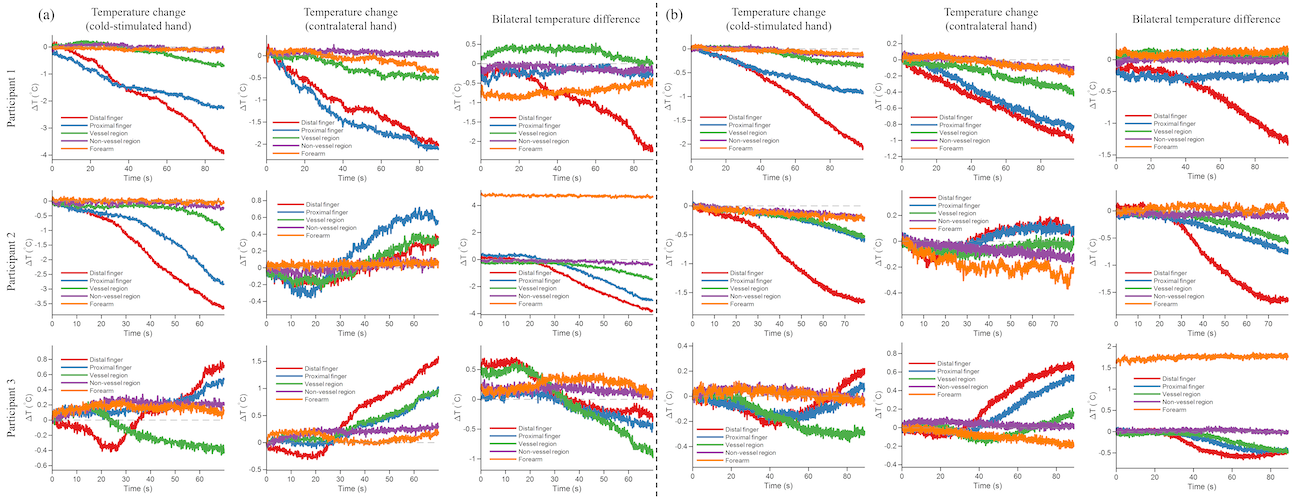}
	\caption{Quantitative thermal-response analysis extracted from five anatomical regions of interest (ROIs) during the contact-cooling experiment. (a) Mid-wave infrared (MWIR) results. (b) Long-wave infrared (LWIR) results. For each participant, the first column shows the temperature change relative to the initial condition, $\Delta T=T-T_0$ for the cold-stimulated hand; the second column presents $\Delta T=T-T_0$ for the contralateral reference hand; and the third column displays the bilateral temperature difference, $\Delta T_{\rm cold-ref}=T_{\rm cold}-T_{\rm ref}$. The curves correspond to five ROIs: distal finger, proximal finger, vessel-associated dorsal-hand region, non-vessel dorsal-hand region, and dorsal forearm. The results reveal spatially heterogeneous cooling responses, with the finger regions generally exhibiting the largest temperature variations and bilateral thermal differences, while the dorsal hand and forearm regions show comparatively weaker responses.}\label{fig6}
\end{figure*}
\section{Numerical Calculation}
To investigate the influence of blood perfusion and thermal contact conditions on the cooling process, a lumped-parameter bioheat model was solved under sustained contact cooling. Figure~\ref{fig1}(a) shows the simulated surface-temperature evolution for different relative perfusion levels. Increasing blood perfusion leads to a progressive reduction in the cooling amplitude because warm arterial blood continuously supplies heat to the tissue. As a result, the skin temperature remains significantly higher at long cooling times for highly perfused conditions. The corresponding baseline-corrected temperature responses shown in Fig.~\ref{fig1}(b) further highlight that enhanced perfusion attenuates the overall temperature decrease throughout the cooling period. The initial cooling rates obtained from linear fitting of the first 15 s of the cooling curves are summarized in Fig.~\ref{fig1}(c). A monotonic decrease in cooling rate is observed with increasing perfusion, indicating that blood flow acts as a thermal buffer that counteracts external heat extraction. Figure~\ref{fig1}(d) presents a sensitivity analysis of the skin–plate contact conductance. Both the initial cooling rate and the final temperature drop increase with increasing contact conductance, demonstrating that efficient thermal coupling between the skin and cooling plate substantially enhances heat removal. Moreover, the initial cooling rate exhibits a nearly linear dependence on contact conductance, whereas the temperature drop increases more gradually at higher conductance values, suggesting a diminishing sensitivity as the system approaches thermal equilibrium. Overall, the numerical results indicate that both blood perfusion and contact conductance strongly influence the transient cooling response, with perfusion primarily regulating heat replenishment and contact conductance controlling the efficiency of heat extraction.

\section{Experimental Setup}
Figure~\ref{fig2} illustrates the experimental configuration used for the bilateral hand contact-cooling measurements. As shown in Fig.~\ref{fig2}(a), one hand was placed on a temperature-controlled cooling plate to induce thermal stimulation, whereas the contralateral hand was positioned on a reference metal plate maintained near room temperature. Three infrared imaging systems operating in different spectral bands, namely long-wave infrared (LWIR), mid-wave infrared (MWIR), and short-wave infrared (SWIR), were arranged above the measurement region to simultaneously monitor the thermal response of both hands. This configuration enabled direct comparison of temperature evolution acquired from different infrared spectral ranges under identical experimental conditions. A photograph of the laboratory setup is presented in Fig.~\ref{fig2}(b), showing the infrared cameras, acquisition computers, display monitors, and metal plates used during the experiments. The synchronized multi-spectral imaging arrangement allowed continuous recording of hand-temperature distributions before, during, and after the contact-cooling procedure, providing the data required for subsequent temporal and spatial thermal analyses.
\begin{figure*}[t]
	\centering
	\includegraphics[width=\textwidth]{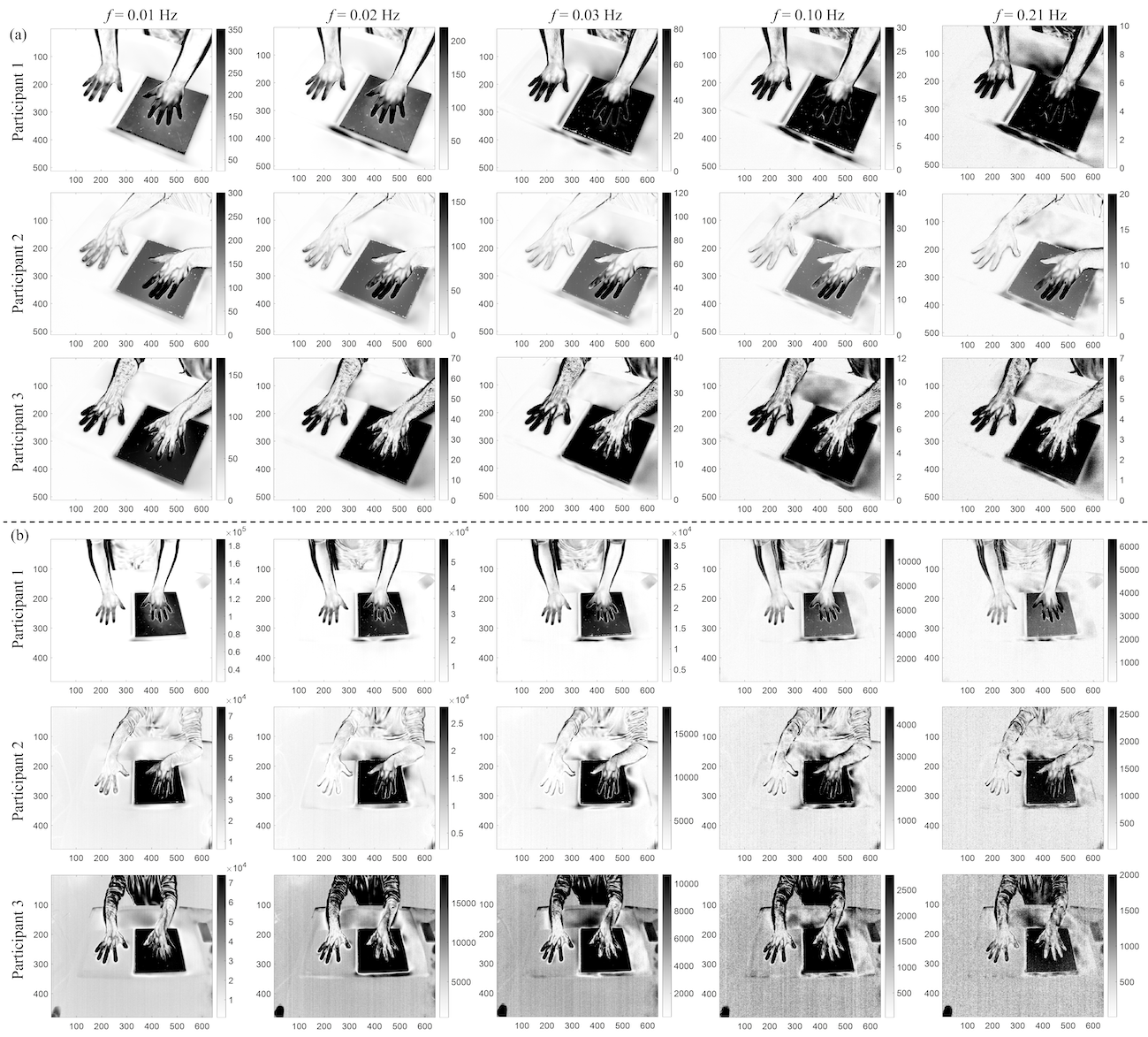}
	\caption{Frequency-domain amplitude maps obtained from thermal image sequences during the contact-cooling experiment. (a) Mid-wave infrared (MWIR) results and (b) long-wave infrared (LWIR) results. For each participant, the amplitude distributions corresponding to the frequency components $f=0.01, 0.02, 0.03, 0.10,$ and 0.21 Hz are shown. The amplitude images were obtained from the discrete Fourier transform of the temporal temperature sequences at each pixel. Low-frequency components exhibit the strongest thermal response and clearly highlight the cooling-induced temperature variations in the hands and forearms, whereas higher-frequency components contain progressively weaker amplitudes and are dominated by localized fluctuations and measurement noise. The results demonstrate that the thermal response associated with contact cooling is primarily concentrated in the low-frequency range.}\label{fig7}
\end{figure*}

\section{Results and Discussion}
Figure~\ref{fig3} presents representative SWIR thermograms acquired during the contact-cooling experiments for the three participants at different time instants. Owing to the limited thermal sensitivity of the SWIR band to near-ambient-temperature objects, the absolute thermal contrast between the hands and the surrounding environment is substantially lower than that observed in the MWIR and LWIR measurements. Nevertheless, the temporal image sequence still reveals a gradual reduction in the apparent temperature of the hand exposed to the cooling plate. The cooled hand becomes progressively darker with increasing contact time, whereas the contralateral hand resting on the reference plate exhibits comparatively smaller variations. Compared with the other infrared spectral bands, the cooling-induced spatial temperature gradients are less pronounced in the SWIR images, and regional thermal details within the fingers and dorsal hand are more difficult to distinguish. These observations indicate that, although SWIR imaging is capable of capturing the overall cooling trend, its sensitivity to subtle skin-temperature changes associated with contact cooling is considerably lower than that of MWIR and LWIR thermography, limiting its effectiveness for quantitative assessment of peripheral thermal responses under the present experimental conditions~\cite{45,46,47,48}.
\begin{table*}[t]
	\caption{
		Initial cooling rates obtained by linear fitting of the early-stage
		temperature responses in five anatomical regions.
		MWIR and LWIR denote the mid-wave and long-wave infrared cameras,
		respectively.
	}
	\label{tab:cooling_rate}
	\centering
	\small
	\setlength{\tabcolsep}{5.5pt}
	\begin{tabular}{
			ll
			S[table-format=+1.4]
			S[table-format=+1.4]
			S[table-format=+1.4]
			S[table-format=+1.4]
			S[table-format=+1.4]
			S[table-format=+1.4]
		}
		\toprule
		\multirow{2}{*}{Condition}
		&
		\multirow{2}{*}{Region}
		&
		\multicolumn{3}{c}{MWIR}
		&
		\multicolumn{3}{c}{LWIR}
		\\
		\cmidrule(lr){3-5}
		\cmidrule(lr){6-8}
		&
		&
		{P1} & {P2} & {P3}
		&
		{P1} & {P2} & {P3}
		\\
		\midrule
		
		\multirow{5}{*}{Cold-stimulated hand}
		& Distal finger
		& 0.0326 & 0.0251 & 0.0119
		& 0.0121 & 0.0103 & 0.0027
		\\
		
		& Proximal finger
		& 0.0344 & 0.0172 & -0.0001
		& 0.0111 & 0.0046 & 0.0042
		\\
		
		& Vessel region
		& -0.0086 & 0.0058 & -0.0066
		& -0.0003 & 0.0055 & -0.0002
		\\
		
		& Non-vessel region
		& -0.0069 & 0.0001 & -0.0102
		& -0.0010 & 0.0038 & -0.0017
		\\
		
		& Forearm
		& 0.0013 & 0.0023 & -0.0062
		& 0.0032 & 0.0027 & -0.0006
		\\
		
		\midrule
		
		\multirow{5}{*}{Contralateral hand}
		& Distal finger
		& 0.0346 & 0.0062 & 0.0098
		& 0.0119 & 0.0072 & 0.0030
		\\
		
		& Proximal finger
		& 0.0482 & 0.0150 & 0.0042
		& 0.0097 & 0.0047 & 0.0009
		\\
		
		& Vessel region
		& 0.0040 & 0.0057 & -0.0017
		& 0.0028 & 0.0036 & 0.0015
		\\
		
		& Non-vessel region
		& -0.0014 & 0.0014 & -0.0108
		& 0.0020 & 0.0019 & -0.0016
		\\
		
		& Forearm
		& -0.0045 & -0.0009 & -0.0033
		& -0.0008 & 0.0079 & 0.0012
		\\
		
		\bottomrule
	\end{tabular}
\end{table*}

Figure~\ref{fig4} compares the temporal thermal responses measured by the MWIR and LWIR cameras during the contact-cooling experiments. In both spectral bands, progressive temperature reductions can be observed in the hand exposed to the cooling plate, whereas the contralateral reference hand exhibits comparatively smaller changes over the same period. The cooling effect becomes increasingly pronounced with contact duration and is particularly evident in the finger regions, which display the largest thermal contrast relative to the surrounding areas. Despite differences in imaging wavelength, the MWIR and LWIR measurements reveal consistent spatial and temporal cooling patterns for all three participants, indicating that the observed thermal response is primarily governed by the underlying heat-transfer process rather than spectral-band-specific effects. Compared with the MWIR images, the LWIR thermograms generally provide higher thermal contrast and clearer visualization of temperature gradients across the hand, especially in the fingers and dorsal hand regions. These results demonstrate that both MWIR and LWIR thermography are capable of capturing the cooling dynamics induced by contact stimulation, while the LWIR measurements offer enhanced sensitivity to subtle surface-temperature variations and improved delineation of spatial thermal features.

\begin{figure*}[t]
	\centering
	\includegraphics[width=\textwidth]{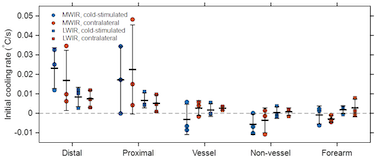}
	\caption{Initial cooling-rate comparison for different anatomical regions obtained from MWIR and LWIR measurements. The initial cooling rates extracted from the distal finger, proximal finger, vessel-associated region, non-vessel region, and forearm are shown for the cold-stimulated hand and the contralateral reference hand. Colored markers represent the values obtained from the three participants, while the black horizontal bars and error bars denote the mean value and one standard deviation, respectively. The finger regions exhibit the highest cooling rates, indicating a stronger thermal response to contact cooling, whereas the vessel-associated, non-vessel, and forearm regions show substantially smaller cooling rates. In most cases, the cold-stimulated hand demonstrates larger cooling rates than the contralateral hand, particularly in the distal and proximal finger regions.}\label{fig8}
\end{figure*}

\begin{figure*}[t]
	\centering
	\includegraphics[width=\textwidth]{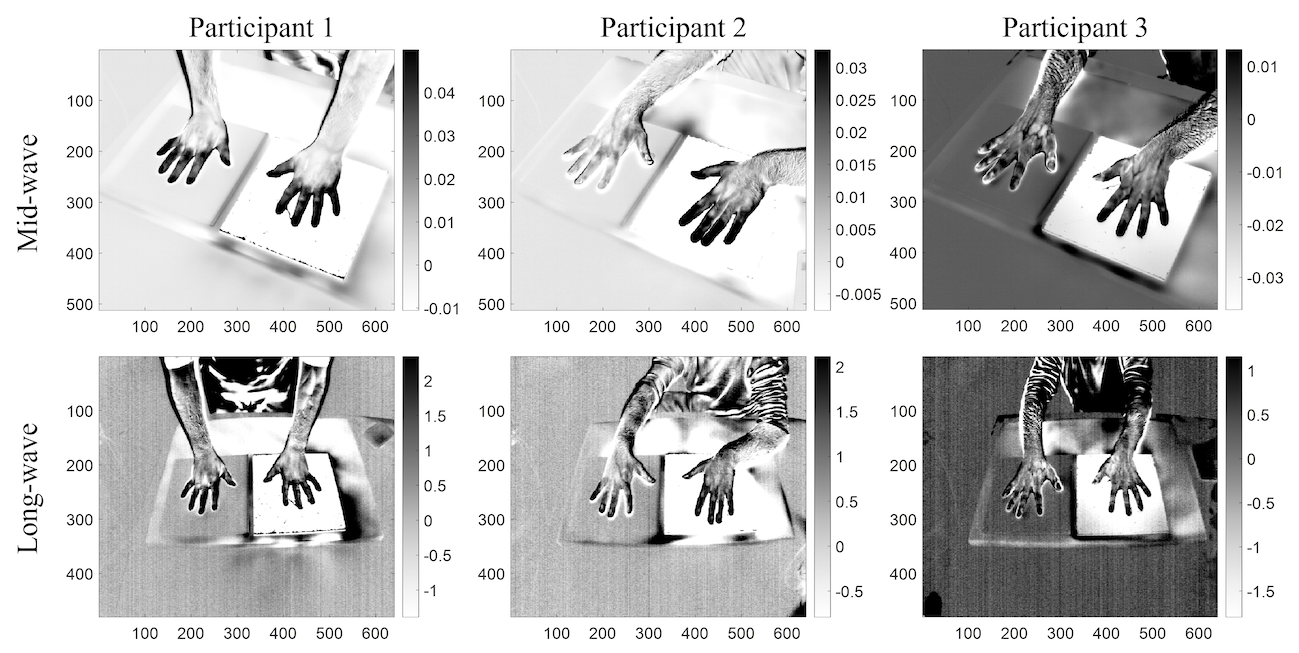}
	\caption{Spatial distribution of the initial cooling rate estimated from pixel-wise temperature responses during the first 20 s of the contact-cooling experiment. Top row: mid-wave infrared (MWIR) results. Bottom row: long-wave infrared (LWIR) results. For each participant, the cooling rate was calculated by applying a linear least-squares fit to the baseline-corrected temperature evolution at every pixel, with the local cooling rate defined as the negative slope of the fitted curve. Larger positive values correspond to faster temperature decreases, whereas values close to zero indicate weak thermal variations. The cooling-rate maps reveal spatially heterogeneous thermal responses, with the strongest cooling generally observed in the hand regions directly exposed to the cooling stimulus, while surrounding areas exhibit substantially weaker cooling dynamics. Differences among participants reflect variations in local heat transfer and physiological thermoregulatory responses.}\label{fig9}
\end{figure*}
Figure~\ref{fig5} presents the temporal temperature evolution extracted from five anatomical regions of interest (ROIs), including the distal finger, proximal finger, vessel-associated dorsal-hand region, non-vessel dorsal-hand region, and dorsal forearm, from both MWIR and LWIR measurements. Overall, the finger regions exhibited the most pronounced thermal response to contact cooling, characterized by a progressive temperature decrease in the cold-stimulated hand and a larger separation from the contralateral reference hand. In contrast, the vessel-associated, non-vessel, and forearm regions showed smaller temperature variations and greater inter-subject variability. Despite differences in spectral range, the MWIR and LWIR results revealed consistent temporal trends across all participants, demonstrating that the observed cooling behavior was reproducible and largely independent of the infrared band. In Participant 1 and Participant 2, the cooled hand generally exhibited lower temperatures than the contralateral hand throughout the experiment, particularly in the distal and proximal finger regions, indicating effective heat extraction by the cooling plate. Participant 3 displayed a different response, with progressive warming trends observed in several regions of both hands, suggesting the influence of physiological factors such as peripheral blood flow regulation and thermal adaptation. Compared with the MWIR measurements, the LWIR data exhibited smoother temperature curves and reduced temporal fluctuations, providing clearer visualization of the cooling dynamics and inter-regional thermal differences. These findings highlight the strong spatial heterogeneity of the hand thermal response and demonstrate that the finger regions are the most sensitive indicators of cooling-induced physiological changes.

Figure~\ref{fig6} presents the baseline-corrected temperature variations ($\Delta T=T-T_0$) and bilateral temperature differences ($\Delta T_{\rm cold-ref}=T_{\rm cold}-T_{\rm ref}$) extracted from the five ROIs using both MWIR and LWIR measurements. For Participants 1 and 2, the cold-stimulated hand exhibited a clear temperature decrease over time, with the largest reductions occurring in the distal and proximal finger regions. This trend was consistently reflected in the bilateral temperature-difference curves, where increasingly negative values indicated a growing thermal contrast between the cooled and reference hands. In contrast, the vessel-associated, non-vessel, and forearm regions generally showed smaller temperature changes and weaker bilateral differences, suggesting lower sensitivity to the external cooling stimulus. Participant 3 displayed a distinct response, characterized by positive temperature variations in several regions of both hands, which may be associated with physiological thermoregulatory adaptation and enhanced peripheral perfusion during the measurement period. Despite the different spectral ranges, the MWIR and LWIR results exhibited remarkably similar temporal behaviors and regional trends, confirming the robustness of the observed cooling dynamics. The bilateral comparison further reduced the influence of baseline temperature variability and common environmental effects, thereby highlighting regional differences in cooling efficiency and vascular thermal regulation across the hand.

Figure~\ref{fig7} presents the frequency-domain amplitude maps obtained from the MWIR and LWIR thermal image sequences at representative frequencies ranging from 0.01 to 0.21 Hz. For both spectral bands, the strongest responses are concentrated at the lowest frequencies, particularly at 0.01–0.03 Hz, where the cooled hand, fingers, and forearm regions are clearly distinguished from the surrounding background. As the frequency increases, the amplitude gradually decreases and the cooling-induced thermal patterns become progressively less pronounced, indicating that the dominant thermal response associated with contact cooling occurs over relatively long temporal scales. This behavior is consistent with the slow heat-transfer processes governed by thermal diffusion and physiological thermoregulation. The MWIR and LWIR amplitude maps exhibit similar spatial distributions across all participants, confirming the reproducibility of the observed frequency-domain features. However, the LWIR results generally present substantially higher signal amplitudes and improved spatial contrast, allowing clearer visualization of the thermal structures associated with the hands and cooling plate. The frequency-domain representation effectively suppresses random temporal fluctuations while emphasizing slowly varying thermal components, demonstrating that most of the information related to contact-cooling dynamics is concentrated in the low-frequency region below approximately 0.05 Hz.

The initial cooling rates obtained from linear fitting of the early-stage temperature responses are summarized in Table~\ref{tab:cooling_rate}. For both MWIR and LWIR measurements, the largest cooling rates were generally observed in the distal and proximal finger regions, indicating that the fingers exhibited the strongest and fastest thermal response to contact cooling. In the cold-stimulated hand, MWIR measurements yielded peak cooling rates of 0.0326–0.0344 $^\circ$C/s for Participant 1 and 0.0172–0.0251 $^\circ$C/s for Participant 2 in the finger regions, whereas substantially lower values were observed in the vessel-associated, non-vessel, and forearm regions. A similar trend was obtained from the LWIR data, although the absolute cooling-rate values were generally smaller. The vessel-associated and non-vessel regions frequently exhibited cooling rates close to zero or slightly negative, suggesting weak temperature variations during the initial stage of the experiment. Moreover, clear inter-subject variability was observed, particularly for Participant 3, whose cooling rates were considerably lower than those of Participants 1 and 2. Despite these quantitative differences, both MWIR and LWIR measurements consistently identified the distal and proximal fingers as the most thermally responsive regions, demonstrating that the initial cooling rate is strongly dependent on anatomical location and may provide a useful indicator for assessing regional thermal and vascular responses to cold stimulation.

Figure~\ref{fig8} compares the initial cooling rates obtained from the MWIR and LWIR measurements for the five anatomical regions. Overall, both infrared modalities produced consistent trends, with the distal and proximal finger regions exhibiting the largest cooling rates, while the vessel-associated region, non-vessel region, and forearm displayed substantially smaller values. The cold-stimulated hand generally showed higher cooling rates than the contralateral hand in the finger regions, reflecting the stronger thermal perturbation induced by contact cooling. MWIR measurements yielded larger cooling-rate magnitudes and a wider dynamic range than LWIR, particularly in the distal and proximal fingers, suggesting greater sensitivity to rapid early-stage temperature changes. In contrast, the LWIR results exhibited lower variability among participants and more clustered cooling-rate values across the different regions. Despite these quantitative differences, both MWIR and LWIR consistently identified the finger regions as the most responsive anatomical locations, while the dorsal hand and forearm demonstrated relatively weak thermal responses. The agreement between the two spectral bands indicates that the observed regional cooling behavior is robust and primarily governed by physiological and heat-transfer processes rather than wavelength-dependent imaging effects.

Figure~\ref{fig9} shows the spatial distribution of the initial cooling rate calculated on a pixel-by-pixel basis from the first 20 s of the temperature evolution for both MWIR and LWIR measurements. In all participants, the highest cooling rates were concentrated in the hand regions directly exposed to the cooling stimulus, particularly in the fingers, where the strongest temperature reductions were observed. The cooling-rate maps reveal substantial spatial heterogeneity, indicating that the thermal response was not uniformly distributed across the hand surface. Regions corresponding to the palm and fingers generally exhibited larger positive cooling rates, whereas the forearm and surrounding background areas showed values close to zero. The MWIR maps provided clear visualization of localized cooling patterns and highlighted regional differences in heat extraction efficiency. Similar spatial trends were observed in the LWIR results, confirming the reproducibility of the cooling-rate distribution across different infrared spectral bands. However, the LWIR maps exhibited a higher level of spatial noise and stronger background texture, whereas the MWIR images provided smoother cooling-rate distributions and clearer delineation of anatomical structures. Inter-subject differences were also evident, with Participants 1 and 2 exhibiting relatively strong cooling responses over the stimulated hand, while Participant 3 showed weaker and more spatially diffuse cooling patterns. These results demonstrate that pixel-wise cooling-rate mapping can effectively identify localized thermal-response characteristics and provide complementary information to ROI-based analyses by revealing the spatial heterogeneity of heat transfer and thermoregulatory activity across the hand.
\section{Conclusion}
In this study, a quantitative thermographic framework was developed to investigate the spatiotemporal thermal response of human hands during controlled contact cooling. Simultaneous measurements were performed using SWIR, MWIR, and LWIR infrared imaging systems, while one hand was exposed to a cooling plate and the contralateral hand served as a reference condition. Temperature evolution was analyzed at multiple anatomical regions, including the distal finger, proximal finger, vessel-associated region, non-vessel region, and forearm. In addition, frequency-domain analysis and initial cooling-rate estimation were applied to characterize the dynamic thermal behavior of the hands. The experimental results demonstrated that the finger regions exhibited the largest temperature reductions and the highest cooling rates, indicating greater sensitivity to external thermal stimulation than the dorsal hand and forearm regions. Bilateral temperature-difference analysis further enhanced the visualization of cooling-induced thermal responses by reducing the influence of baseline temperature variability and common environmental effects. MWIR and LWIR measurements produced highly consistent temporal and spatial trends, while LWIR images generally provided higher thermal contrast and improved sensitivity to subtle surface-temperature variations. In contrast, the SWIR measurements were capable of capturing the overall cooling process but exhibited lower thermal contrast under near-ambient conditions.
Frequency-domain analysis revealed that the dominant thermal response was concentrated at low frequencies below approximately 0.05 Hz, corresponding to the slow thermal dynamics associated with conductive heat transfer and physiological thermoregulation. Pixel-wise cooling-rate maps further demonstrated substantial spatial heterogeneity across the hand surface and provided a detailed visualization of regional cooling efficiency. The numerical bioheat model showed that both blood perfusion and skin–plate contact conductance strongly influence the cooling behavior, with perfusion acting as a thermal buffering mechanism and contact conductance governing the efficiency of heat extraction.
Overall, the results demonstrate that dynamic infrared thermography combined with cooling-rate analysis, bilateral comparison, and frequency-domain processing provides a robust approach for assessing regional thermal responses of the hand during controlled contact cooling. The proposed methodology may serve as a foundation for future studies involving larger subject populations and may contribute to the development of non-contact techniques for evaluating peripheral circulation, vascular regulation, and thermoregulatory dysfunction.

\begin{acknowledgments}
This work was supported by the Natural Sciences and Engineering Research Council (NSERC) Canada through the CREATE 'oN DuTy!' program (No. 496439-2017) and the Canada Research Chair in Multi-polar Infrared Vision (MiviM).
\end{acknowledgments}

\appendix



\end{document}